\documentclass{ringb99}
\usepackage{graphics}

\begin{document}
\title{The heating of the ICM by powerful radio sources}
\author{C. R. Kaiser\inst{1} \and P. Alexander\inst{2}}  
\institute{Max--Planck--Institut f\"ur Astrophysik,
Karl-Schwarzschild-Str.1, 85740 Garching, Germany
\and  Cavendish Laboratory, Madingley Road, Cambridge, CB3 0HE, UK}
\maketitle

\begin{abstract}
We present a model for the compression and heating of the ICM by
powerful radio galaxies and quasars. Based on a self-similar model of
the dynamical evolution of FRII-type objects we numerically integrate
the hydrodynamic equations governing the flow of the shocked ICM in
between the bow shock and the radio lobes of these sources. The
resulting gas properties are presented and discussed. The X-ray
emission of the shocked gas is calculated and is found to be in
agreement with observations. The enhancement of the X-ray emission of
cluster gas due to the presence of powerful radio galaxies may play an
important role in the direct detection of cluster gas at high
redshifts. 
\end{abstract}

\section{Introduction}

Since jets were identified as the means by which AGN transport their
energy to the large scale structure of radio galaxies and radio loud
quasars observed at radio wavelengths (Rees 1971\nocite{mr71}, Scheuer
1974\nocite{ps74}), much effort has been devoted to the question how
the environments of the hosts of extragalactic radio sources influence
the evolution of these objects. On the other hand, the influence of
powerful radio sources on the properties and the evolution of their
environments is still largely unknown. This is partly caused by the
difficulty involved in observing the interaction of the large scale
structure of radio sources with its surroundings. 

In the case of the lower luminosity version of radio galaxies (type
FRI, Fanaroff \& Riley 1974\nocite{fr74}) high resolution radio maps
show a very complex and turbulent jet flow. This turbulence is caused
by the passage of the jet through the ICM. In the case of the more
powerful version (FRII) the interaction with the ICM is not so
directly visible. In these objects the jets are much more collimated
and end at distinct points; the radio hot spots. The lobe-like radio
structure of FRII-type sources surrounding the jets is thought to
expand supersonically with respect to the ICM. However, the resulting
bow shock is inferred theoretically but has been detected indirectly in
only one source (Carilli, Perley \& Dreher 1988\nocite{cpd88}).

Despite this lack of directly observable signatures of the influence
of radio sources on their environments, the resulting effects on the
properties and the evolution of the ICM should be quite
significant. The confinement of the large scale structure of radio
sources of type FRII implies that a considerable fraction of the
energy transported by the jets is transfered to the ICM. The shocks
around the radio structures of FRII-type objects are presumably very
effective at converting the expansion energy of the radio lobes into
thermal energy of the ICM. Furthermore, the jets in FRIIs are found to
be more powerful than those in FRI-type sources (e.g. Rawlings \&
Saunders 1991\nocite{rs91}). This means that the `stronger flavour'
FRII radio sources may also influence the ICM more strongly.

At low redshift FRIIs are predominantly found in poor groups and
rarely in clusters. However, there is some evidence that at high
redshift these powerful sources are located in richer environments
(e.g. Hill \& Lilly 1991\nocite{hl91}). This may imply that at earlier
cosmological epochs FRII-type radio sources were more common in
proto-cluster environments possibly influencing their evolution
significantly. 

In this paper we outline a model describing the interaction of
FRII-type radio sources with their environments and the resulting
enhancement of the X-ray emission of this material. This represents a
first step towards unraveling the complex relation between the
evolution of powerful active galaxies in clusters and that of their
surroundings.

\section{Analytical model of the evolution of FRII radio sources}

It is obviously not possible to observe one source at various stages
of its life thereby `watching' the evolution of a given radio
source. However, complete samples of radio galaxies and quasars
provide us with a view of many sources of very different ages. If the
evolution of these sources is not completely different from one
another than we can hope to reconstruct some of the characteristics of
this evolution. Leahy \& Williams (1984)\nocite{lw84} and Leahy,
Muxlow \& Stephens (1989)\nocite{lms89} find a correlation between the
radio luminosity of FRII radio galaxies of linear sizes up to 500 kpc
with the aspect ratio of their radio lobes, i.e. the ratio of the lobe
widths to their lengths, $R$, but find no correlation of $R$ with the
linear sizes of the radio lobes. Subrahmanyan, Saripalli \& Hunstead
(1996)\nocite{ssh96} find values of $R$ for sources exceeding 900 kpc
in linear size which are very similar to the ones found for the
smaller sources. This strongly suggests that the large scale structure
of radio galaxies and radio loud quasars with FRII morphology grows in
a self-similar fashion. Any theoretical model of the development of
these structures must therefore comply to this observational
constraint.

Scheuer (1974)\nocite{ps74} and Begelman \& Cioffi (1989)\nocite{bc89}
argued that the region of the radio lobes or cocoon should be
overpressured with respect to the surrounding gas and will therefore
drive a strong shock, the bow shock, into this material. Falle
(1991)\nocite{sf91} then showed that the expansion of this bow shock
should be self-similar once the mass of the ICM swept up by this shock
and pushed aside by the cocoon exceeds the mass of the material within
the cocoon. This requirement is fulfilled for virtually any radio
galaxy of a linear size exceeding a few kpc. In Kaiser \& Alexander
(1997)\nocite{ka96b} we developed a model which predicts the
self-similar growth of FRII-type radio galaxies. The following is a
brief summary of this model and its assumptions. 

We assume that the two jets emerge symmetrically to both sides from
the AGN in the centre of the host galaxy. The jets end in strong
shocks. These shocks and their immediate surroundings form the
observed radio hot spots. The jet material then inflates the cocoon
which shields the jets from the denser material of the ICM. Because of
the very high sound speeds within this post-shock material the
pressure in the cocoon will equalize rapidly and so the pressure
within the cocoon is uniform throughout with the exception of the hot
spot region. The magnetic field in the cocoon is assumed to be tangled
on scales smaller than any of the dynamically relevant scales. It
therefore contributes to the overall pressure within the cocoon but
does not exert any non-isotropic stresses. This implies that the jets
are confined solely by the pressure within the cocoon. The source is
embedded in the ICM and we model the gas density distribution within
this material with a power law. This is a reasonable approximation at
large radii to the more realistic King (1972)\nocite{ik72} profile
which provides good fits to the observed X-ray brightness profiles in
the vicinity of radio galaxies and that of clusters. The cocoon
expands into this material at a velocity exceeding the local sound
speed. The whole radio source is therefore surrounded by a bow shock.

Using these simple assumptions we find that not only the growth of the
bow shock but also that of the cocoon, i.e. the radio lobes, should be
self-similar. Note that this is a result of the model and not an
assumption. Furthermore, to arrive at this result we have made no
assumptions about the shape of the bow shock or the cocoon. With this
model we calculate that about half of the energy transported by a
`typical' jet of a radio source of type FRII is eventually converted
to kinetic and thermal energy of the shocked ICM located between the
bow shock and the boundary of the cocoon. The jets of a conventional
radio galaxy have an energy transport rate of roughly $10^{46}$ ergs
s$^{-1}$. This implies that such an object deposits about $10^{61}$
ergs of energy in the ICM during its lifetime of $10^8$ years. This
enormous amount of energy will certainly influence the subsequent
evolution of any gaseous structure the radio galaxy is embedded in.

\section{Shock heating of the ICM}

Because we do not make any assumptions about the shape of the bow
shock and the cocoon in the model described in the previous Section,
it is impossible to determine from this model alone how and in which
form the energy transfered from the jet to the shocked ICM is
distributed. The properties of the gas in the layer in between the bow
shock and the boundary of the cocoon are essentially described by the
usual hydrodynamical equations; the equations of conservation of mass,
momentum and energy. The problem resembles that of the gas flow behind
the shock front of a strong explosion presented by Sedov and Taylor
(e.g. Sedov 1959\nocite{ls59}). Dyson, Falle \& Perry
(1980)\nocite{dfp80} present an extension to the explosion case in the
form of a spherically symmetric wind with constant energy input
driving the shock front. In both cases the solutions are
self-similar. This and the results of our own model described in the
previous Section encouraged us to construct a model for the layer of
shocked gas surrounding the cocoons of FRII-type radio sources. The
main difference to the previous work is the elongated geometry of the
radio source and the therefore more complex structure of the
hydrodynamical equations. The method and results presented in the
following are described in more detail in Kaiser \& Alexander
(1999)\nocite{ka98b}.

In their simplest form the conservation equations present a problem
with four independent variables; three spatial and one time
variables. To reduce the complexity we assume an adiabatic equation of
state for the shocked gas and axisymmetry about the jet axis. The
self-similar expansion of the cocoon and the bow shock then allows us
to recast the equations in a co-moving coordinate system which expands
with the bow shock and the boundary of the cocoon. In this system
these two surfaces are at rest. This then reduces the number of
independent variables to two spatial coordinates. Our dynamical model
predicts the growth of the radio source to be self-similar independent
of the exact geometrical shape of the bow shock and that of the
cocoon. We are therefore free to choose a shape for either of these
surfaces to fit observations. Once we have chosen a specific shape for
one of these surfaces and the initial conditions along this starting
surface, it is possible to numerically integrate the hydrodynamical
equations in the direction of the other surface until it is reached by
the solution. Although it seems natural to choose the boundary of the
cocoon to start the integration from, since the cocoon is driving the
bow shock, it is more appropriate to start from the bow shock. This is
because in the co-moving coordinate system the gas is apparently
streaming through the bow shock into the shocked region towards the
cocoon boundary. Also, if we assume the bow shock to be strong than
the initial conditions of the gas just behind the bow shock are well
determined. For the shape of the bow shock and therefore for that of
the co-moving coordinate system we choose a prolate spheroid. The
resulting shape of the cocoon boundary found from the numerical
integration is very similar and therefore agrees well with
observations of radio lobes. For details of the calculation and the
numerical method we refer the reader to Kaiser \& Alexander
(1999)\nocite{ka98b}.

\section{Properties of the shocked ICM}

\begin{figure*}
\resizebox{12cm}{!}{\includegraphics{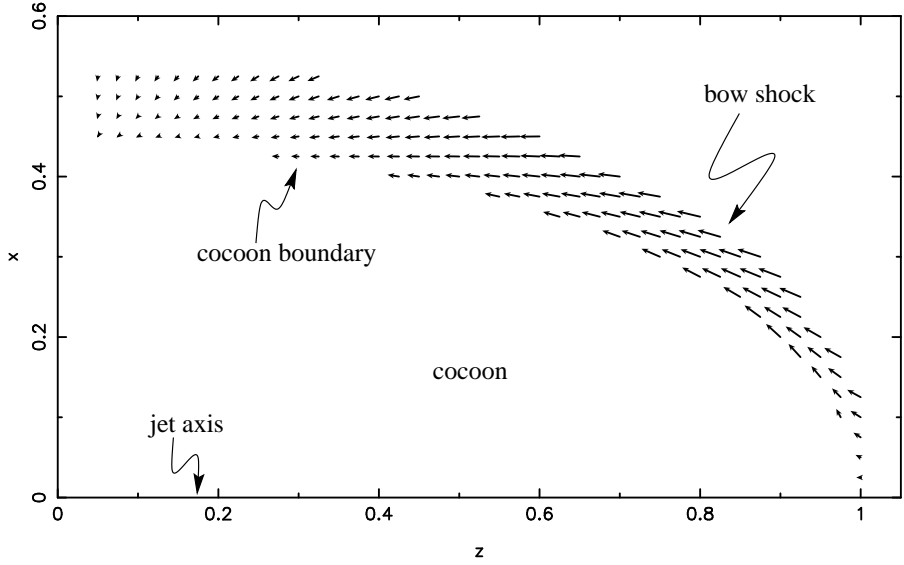}}
\caption[]{The velocity field of the gas in the shocked layer in
between bow shock and cocoon. In this and the other figures we only
show one quarter of the radio source. Because of the assumed
rotational symmetry about the jet axis all velocity vectors are not
projected and lie fully in the plane of the plot. The centre of the
host galaxy is located at the origin of the plot while the jet ends
roughly at x=0 and z=1. The velocities are shown in the rest frame of
the cocoon boundary and the bow shock. The axes are in units of the
cocoon length.}
\label{fig:vel}
\end{figure*}

Figure \ref{fig:vel} shows the velocity field of the gas within the
shocked layer. The gas is flowing along the boundary of the
cocoon. The flow lines are bent parallel to this surface. The two
stagnation points in front of the end of the jet and close to the
x-axis away from the bow shock are caused by the symmetrical boundary
conditions which do not allow any gas flow across the jet axis or
through the surface defined by the rotation of the x-axis about the
jet axis into the other half of the cocoon. The cocoon in Figure
\ref{fig:vel} is relatively broad. This is for illustrative purposes
only. The flow patterns in `slimmer' cocoons are very similar.

\begin{figure*}
\resizebox{14cm}{!}{\includegraphics{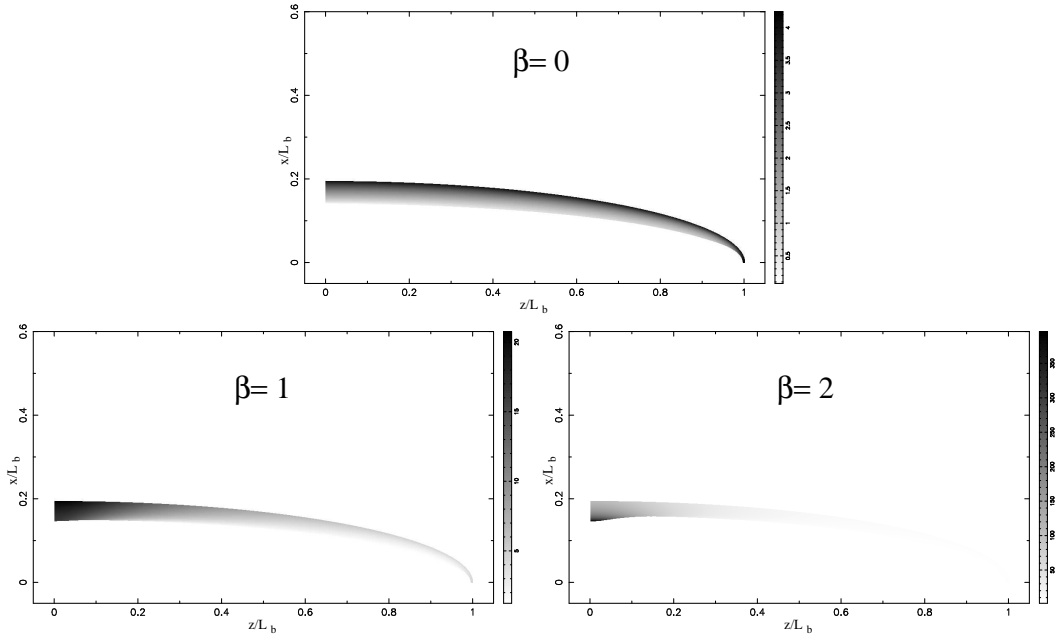}}
\caption[]{The density distribution of the shocked gas in between the
bow shock and the cocoon boundary. $\beta$ is the exponent of the
power law describing the density profile of the unshocked ICM in front
of the bow shock.}
\label{fig:den}
\end{figure*}

The distribution of the gas density within the shocked layer is shown
in Figure \ref{fig:den} for three different density profiles of the
unshocked ICM in front of the bow shock. The case of a uniform
density of the ICM is qualitatively similar to the findings of Sedov
(1959)\nocite{ls59} and Dyson et al. (1980)\nocite{dfp80}. The density
decreases away from the bow shock towards the cocoon boundary. This is
the signature of the gas that is re-expanding after being compressed by
the bow shock. This general trend is much weaker in the case of $\beta
=1$ and is reversed for $\beta =2$. In the latter case the expansion
of the gas behind the bow shock is so slow that the original density
distribution, i.e. decreasing density with increasing distance from
the source centre, is preserved. This gives the impression of a
further compression of the gas on its way from the bow shock to the
cocoon boundary. It is interesting to note that this reversal of the
density profile in the shocked gas is not a result of the elongated
geometry of the bow shock but is also present in the spherical case
(see Kaiser \& Alexander 1999\nocite{ka98b}).

\begin{figure*}
\resizebox{14cm}{!}{\includegraphics{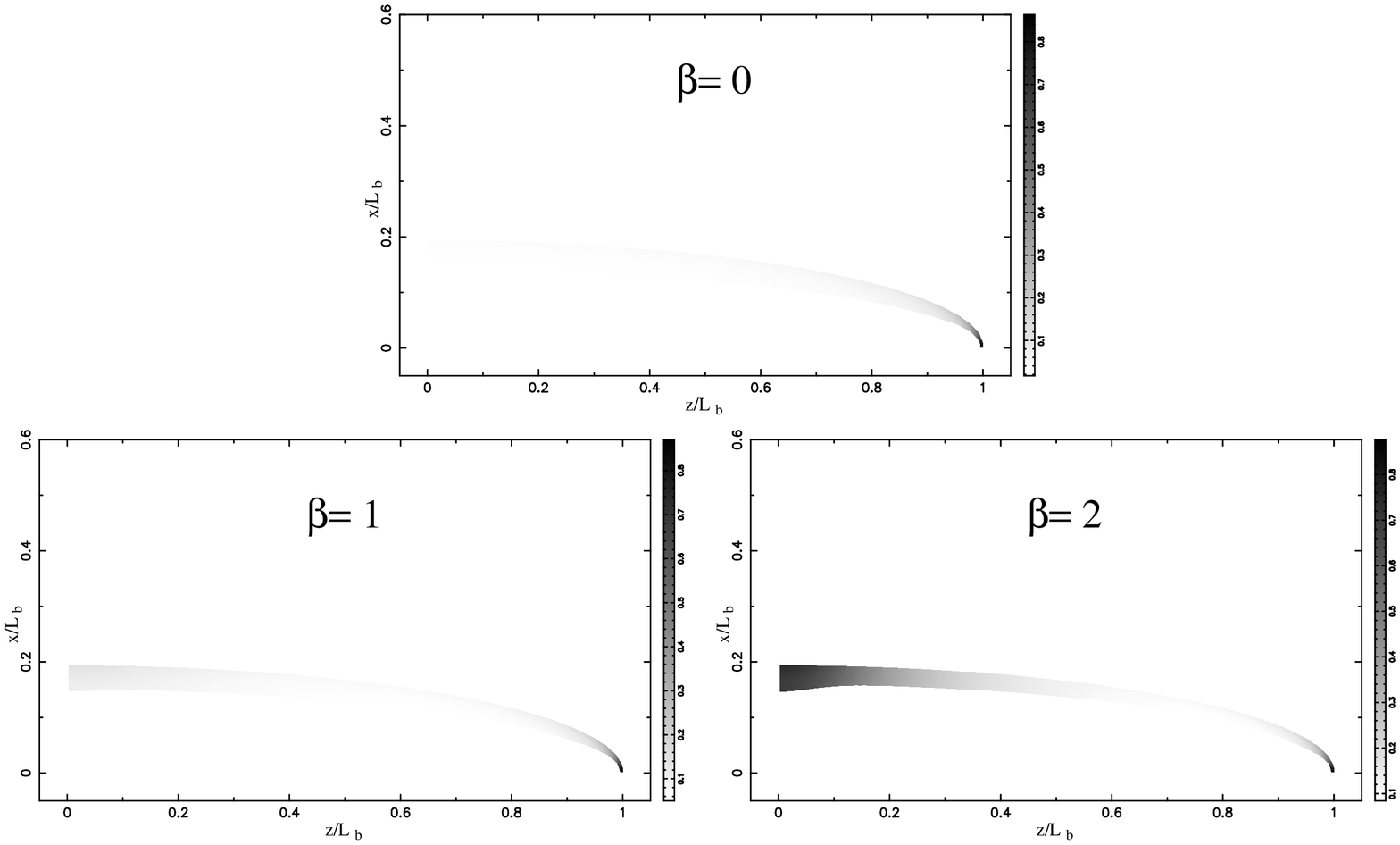}}
\caption[]{The pressure distribution of the shocked gas in between bow
shock and cocoon boundary.} 
\label{fig:pre}
\end{figure*}

The pressure distribution within the shocked gas is presented in
Figure \ref{fig:pre}. Away from the region immideatly in front of the
jet end the pressure is decreasing away from the bow shock towards the
cocoon for $\beta =0$ and $\beta =1$. This is the behaviour found by
Dyson et al. (1980)\nocite{dfp80} in the spherical case. However, in
front of the hot spot where the jet ends the pressure is increasing
towards the cocoon for all three cases. We find the same pattern all
along the bow shock for $\beta =2$. The pressure at the cocoon
boundary also varies along this surface. In the case of a uniform
density of the unshocked ICM we find a monotonic decrease of the
pressure from the hot spot region towards the x-axis. For the other
cases studied here the pressure also decreases away from the hot spot
but then rises again towards the x-axis. In the case of $\beta =2$ the
pressure at the cocoon boundary on the x-axis is almost as high as in
front of the hot spot. This also causes the slight bend in the cocoon
surface towards the jet axis in this case. The pressure at the cocoon
boundary must be matched by the material inside the cocoon. The strong
variation of the pressure along this surface in the case $\beta =2$
therefore clearly violates our assumption of a uniform pressure within
the cocoon. This inconsistency may imply that the cocoons of FRII
sources embedded in environments with steep density gradients do not
grow self-similarly or that the shape of the bow shocks in front of
them is not well approximated by the prolate spheroid we have assumed
here. In the second case significant off-axis flow of the radio plasma
within the cocoon may occur. Note also that we have assumed an
adiabatic equation of state for the shocked gas. In the high pressure
- high density region close to the x-axis of sources embedded in
environments with steep density gradients this approximation may not
be appropriate and radiative cooling could be very effective. This
would decrease the pressure in these regions and may lead to a
self-consistent model also for steep external density gradients.

We have assumed that the density distribution of the unshocked ICM
follows a power law. This is a good approximation to the more
appropriate King (1972)\nocite{ik72} profile at large radii. The King
profile corresponds roughly to the equilibrium distribution of an
isothermal sphere of gas. We therefore expect the temperature of the
shocked gas not too vary strongly with position in the flow since the
initial conditions behind the bow shock are not a very strong function
of position either. Our calculations show that there are indeed no
large temperature fluctuation within the layer of shocked gas. The
only exception is the region in front of the hot spot where the
temperature is somewhat higher than in the rest of the gas.

\begin{figure*}
\resizebox{14cm}{!}{\includegraphics{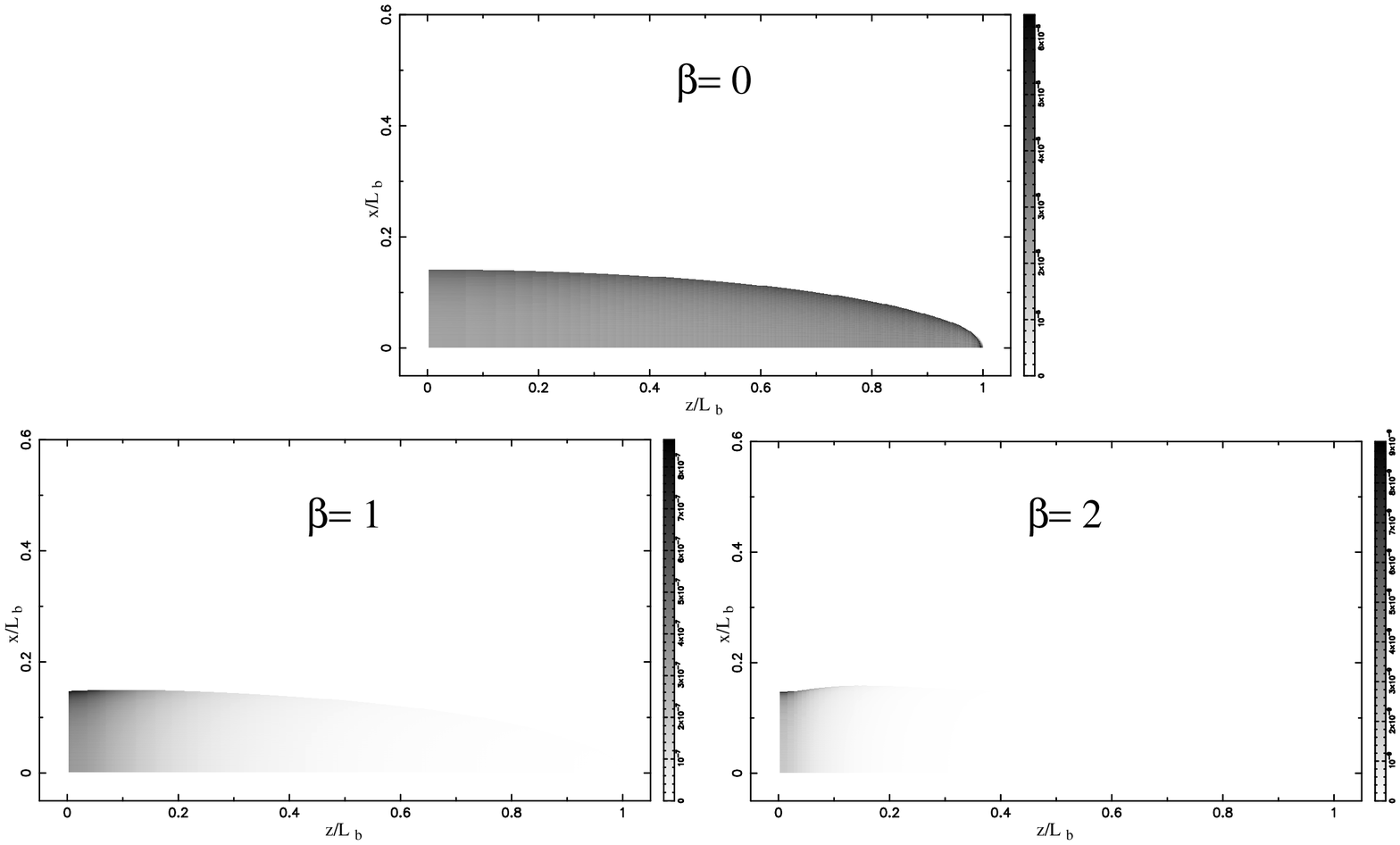}}
\caption[]{The X-ray surface brightness in the 0.1 to 2 keV band.}
\label{fig:bri}
\end{figure*}

Since the density and the temperature of the shocked gas is known from
the model, it is straightforward to calculate the X-ray surface
brightness of the shocked gas layer due to thermal bremsstrahlung. For
this we used the emission of the gas in the range from 0.1 to 2 keV,
appropriate for the High Resolution Imager (HRI) on board ROSAT. We
rotated the source about the jet axis and then calculated the surface
brightness by integrating along line of sights perpendicular to the
plane of the map shown here. This of course assumes that the source
lies in the plane of the sky. In Figure \ref{fig:bri} we show the
results of this calculation for the three cases discussed here.

For a steep external density gradient the X-ray emission is very
concentrated just behind the bow shock close to the x-axis. This can
be explained with the high gas density in this region. In general the
X-ray surface brightness is predicted by our model to be a good tracer
of gas density. However, this of course is mainly due to the
assumption of an isothermal sphere for the unshocked gas. If
temperature fluctuations are present in the unshocked ICM, they will
disturb the rather smooth result presented here. 

The only FRII-type radio galaxy which has been observed with enough
spatial resolution and sensitivity to detect this emission from the
shocked ICM to date is Cygnus A. The HRI map presented by Carilli,
Perley \& Harris (1994)\nocite{cph94} shows an enhancement of the
X-ray emission in the correct place; along the radio lobes close to
the core of the host galaxy. A more quantitative analysis of the X-ray
map in this case is difficult because the presence of the radio source
precludes an accurate determination of the properties of the unshocked
IGM. However, using the model outlined above we find some evidence
that the gas causing the mentioned X-ray emission peaks may have been
denser than expected at this position from a pure King profile before
it passed through the bow shock. Additional radiative cooling in these
regions as outlined above may also contribute to the higher density
here.

Another intriguing example for a possible detection of enhanced X-ray
emission of shocked gas due to the effects of jets is the HRI map of
the possible cluster around the powerful radio galaxy 3C 356 by
Crawford \& Fabian (1996)\nocite{cf96}. The distance of this object
(z=1.079) is far too great to allow an observation as well resolved as
the one of Cygnus A. However, the possible extension of the X-ray
detection is well aligned with the axis of the radio source as we
would expect from our model. The heating of the ICM by powerful radio
sources together with the scattering of nuclear emission from the AGN
(Setti, these proceedings) may therefore play an important role in the
direct detection of high redshift clusters because both processes
enhance the X-ray emission of the cluster gas.

\section{Summary}

We have presented an analytical model for the dynamical evolution of
powerful extragalactic radio sources of type FRII. This model predicts
self-similar growth of the bow shock and radio lobes/cocoon of the
sources independent of the exact geometrical shape of these
surfaces. Based on this model we performed a numerical integration of
the hydrodynamical equations governing the gas flow of the shocked ICM
between the bow shock and the boundary of the cocoon of FRII
sources. We find that the X-ray surface brightness of the shocked gas
is boosted significantly by the presence of the shock. In the case of
an initially isothermal density distribution of the ICM the enhanced
X-ray surface brightness distribution is a good tracer of gas density
within the shocked layer of gas. The model is in agreement with the
high resolution X-ray observations of Cygnus A. 

The presence of a powerful radio source in a cluster clearly boosts
the X-ray emission of the ICM. This may help us in detecting this
material at high redshifts. The influence of radio sources on the
evolution of the ICM still remains to be investigated in detail. The
study presented here is a first step towards this goal.

\bibliography{../../crk}
\bibliographystyle{../../mnras}

\end{document}